\documentclass[nonacm,sigconf,screen]{acmart}

\renewcommand\footnotetextcopyrightpermission[1]{}
\acmConference[]{}{}{}
\AtBeginDocument{%
  \providecommand\BibTeX{{%
    \normalfont B\kern-0.5em{\scshape i\kern-0.25em b}\kern-0.8em\TeX}}}

\usepackage{amsmath}
\usepackage{pgfplots}
\pgfplotsset{compat=1.17}
\usepackage{caption}
\usepackage{subcaption}
\usepackage{graphicx}
\usepackage{xparse} 
\usepackage{xspace}
\usepackage{wrapfig}
\usepackage{comment}
\usepackage{multirow}
\usepackage{lipsum}
\usepackage{rotating}
\usepackage{enumitem}
\usepackage{float}
\usepackage{mdframed} 

\newtoggle{showmarks}
\toggletrue{showmarks}  
\newcommand{\CompanyX}{Microsoft}

\AtEndPreamble{
    \usepackage[capitalize]{cleveref}
    \crefname{section}{Sec.}{Secs.}
    \Crefname{section}{Section}{Sections}
    \crefname{table}{Tab.}{Tabs.}
    \Crefname{table}{Table}{Tables}
    \crefname{figure}{Fig.}{Figs.}
    \Crefname{figure}{Figure}{Figures}
}

\makeatletter
\DeclareRobustCommand\onedot{\futurelet\@let@token\@onedot}
\def\@onedot{\ifx\@let@token.\else.\null\fi\xspace}

\def\eg{\emph{e.g}\onedot} 
\def\ie{\emph{i.e}\onedot} 
 
\def\etc{\emph{etc}\onedot}

\def\etal{\emph{et al}\onedot}
\makeatother

\setlist[itemize]{leftmargin=*, topsep=2.5pt}
\setlist[enumerate]{leftmargin=*, topsep=2.5pt}

\newcommand{\ffbox}[1]{
    \begin{mdframed}
    \small%
    #1
    \end{mdframed}
}
\definecolor{grayish}{HTML}{adb5bd}
\NewDocumentCommand{\glipsum}{O{1} O{}}{%
  \ifx\relax#2\relax
    {\color{grayish}\lipsum[#1]}%
  \else
    {\color{grayish}\lipsum[#1][1-#2]}%
  \fi
}
\renewcommand{\paragraph}[1]{\vspace{1mm}\noindent{#1}}

\definecolor{ak}{HTML}{00b4d8}
\definecolor{dg}{HTML}{a47e1b}
\definecolor{fz}{HTML}{bd68ee}
\definecolor{ap}{HTML}{37B5B6}
\definecolor{sg}{HTML}{38b000}

\setcopyright{acmcopyright}
\copyrightyear{2024}
\acmYear{2024}
\acmDOI{XXXXXXX.XXXXXXX}

\acmPrice{15.00}
\acmISBN{978-1-4503-XXXX-X/18/06}

\setcopyright{none}
\renewcommand\footnotetextcopyrightpermission[1]{} 
\renewcommand{\shortauthors}{Goel et al.}

\begin{document}

\title{X-lifecycle Learning for Cloud Incident Management using LLMs}
\author{Drishti Goel, Fiza Husain, Aditya Singh, Supriyo Ghosh, Anjaly Parayil, Chetan Bansal,\\  Xuchao Zhang, Saravan Rajmohan}
\email{{t-drgoel, t-fizahusain, t-aditysingh, supriyoghosh, aparayil, chetanb, xuchaozhang, saravan.rajmohan}@microsoft.com}
\affiliation{%
   \institution{Microsoft}
  \streetaddress{}
   \city{}
   \country{}
}


\renewcommand{\shortauthors}{}

\begin{abstract}
Incident management for large cloud services is a complex and tedious process and requires significant amount of manual efforts from on-call engineers (OCEs). OCEs typically leverage data from different stages of the software development lifecycle [SDLC] (\eg, codes, configuration, monitor data, service properties, service dependencies, trouble-shooting documents, \etc) to generate insights for detection, root causing and mitigating of incidents. Recent advancements in large language models [LLMs] (\eg, ChatGPT, GPT-4, Gemini) created opportunities to automatically generate contextual recommendations to the OCEs assisting them to quickly identify and mitigate critical issues. However, existing research typically takes a silo-ed view for solving a certain task in incident management by leveraging data from a single stage of SDLC. In this paper, we demonstrate that augmenting additional contextual data from different stages of SDLC improves the performance of two critically important and practically challenging tasks: (1) automatically generating root cause recommendations for dependency failure related incidents, and (2) identifying ontology of service monitors used for automatically detecting incidents. By leveraging 353 incident and 260 monitor dataset from \CompanyX, we demonstrate that augmenting contextual information from different stages of the SDLC improves the performance over State-of-The-Art methods. 
\end{abstract}

\begin{comment}

\end{comment}





\maketitle

\section{Introduction} \label{intro} 

Large-scale cloud service providers such as Microsoft, Google, Amazon, Salesforce run tens of thousands of services continuously in a complex and distributed environment. Despite significant efforts and best practices, cloud incidents (\eg, performance degradation or unplanned interruptions) are inevitable and can be caused from failure or errors from any stage of the software development lifecycle [SDLC] (\eg, code or configuration errors, dependency failures, monitor logic malfunction, hardware issues, \etc). These incidents can be remarkably expensive for the organization in terms of customer dissatisfaction, economical penalties and engineering manpower required to resolve the incident. For instance, during a major event or festival season, the estimated cost for one hour of service downtime for Amazon is approximately US\$100 million \cite{amazon-100-million}.  

Artificial Intelligence for IT Operations (AIOPs) has gained popularity where data-driven AI techniques are used to automate parts of the incident life-cycle including incident prioritization~\cite{chen2020incidental}, similar incident linking~\cite{chen2020identifying,ghosh2024dilink}, and reducing the time to mitigate (TTM) incidents~\cite{chen2019continuous,jiang2020mitigate}.
Motivated by recent advancements in large language models (LLMs) (\eg, ChatGPT~\cite{openai2023-chatgpt}, GPT-4~\cite{Achiam2023GPT4TR}, Gemini~\cite{Anil2023GeminiAF}), there has been a increasing momentum in harnessing LLMs to automate various aspects of the incident life cycle. These applications span from intelligent monitoring for detection~\cite{ganatra2023detection,Srinivas2024IntelligentMonitoring} to automated query recommendations~\cite{jiang2023xpert} and automated root causing and mitigation~\cite{ahmed2023recommending, zhang2024automated, chen2023empowering} of incidents. The existing research often focuses on leveraging contextual information relevant to individual incidents or services, thereby overlooking the complex interactions and data communication across different stages of the software development life cycle (SDLC). However, in practice, on-call engineers (OCEs) takes into consideration a wide variety of cross-lifecycle (X-lifecycle) contextual information for understanding and resolving incidents. In this work, we identified two such practically important and challenging scenarios:
\begin{enumerate}
    \item \textbf{Root cause analysis for dependency failures:} Diagnosing incident typically requires significant manual effort and back-and-forth communication among OCEs. Existing research only leverage incident meta-data (\eg, incident title and summary) for either finetuning LLMs~\cite{ahmed2023recommending} or directly querying pretrained LLMs with in-context examples~\cite{zhang2024automated, chen2023empowering}. However, for certain categories of incidents such as failures stemming from dependent partner services, we need to consider additional information (\eg, upstream service dependencies and their functionalities) for holistic understanding and reasoning.
    \item \textbf{Monitor categorization:} Automatically learning a structured ontology for monitors is useful for developing data-driven intelligent monitoring systems. \cite{Srinivas2024IntelligentMonitoring} leverage LLMs using only monitor metadata to categorize monitors based on what to monitor (resource class) and which metrics to monitor (SLO class). 
    However, there are no specific templates for the monitor metadata, and it highly depends on the engineer who creates it. As a result, it often includes a lot of software, deployment environment, and service-specific information that may not provide sufficient context on what exactly is being monitored (resource class) and which metrics are being monitored (SLO class). Therefore, additional information on the service and component functionality needs to be captured for additional context.
\end{enumerate}

In this work, we propose to leverage X-lifecycle data (\eg service architecture, dependencies and functionalities) for enhancing the quality of automated recommendations from LLMs for both root cause analysis of dependency failures and monitor categorization problem.  
Our goal is to answer the following research questions: (a) \textbf{RQ1}: In a cloud setting with distributed architecture, how does data from different stages of the service lifecycle influence decision making and how can we propose approaches to leverage this data for better decision-making? (b) \textbf{RQ2}: How upstream service dependencies and their service property information help in root cause analysis of dependency failures? (c) \textbf{RQ3}: How does the influence of service architecture and service functionality vary in identifying SLO and resource categories of the monitor?


To answer these questions, we conducted extensive experiments on real-world incident and monitor dataset from \CompanyX.
For root cause recommendation of dependency failures, we identified 353 (roughly 50\% of these are dependency failures) high impact historical incidents from Intelligent Conversation and Communication Cloud (IC3) service which supports the back-end of Teams application, serving more than 250 million users worldwide. We first summarize the root causes and incident summaries to reduce noise and input token length.
Motivated by Zhang \emph{et al.}\cite{zhang2024automated}, we create an end-to-end pipeline to retrieve top 5 similar incidents from historical data corpus, which are used as in-context examples.
In addition to incident metadata, we curate the upstream service dependencies along with their service properties and functionalities for a given service.
Finally, for a given incident originated from a particular service, we generate a prompt with in-context examples and upstream service descriptions and query GPT-4 model to generate the plausible root cause information.
Using human evaluation along with lexical and semantic evaluation metrics, we demonstrate that augmenting service properties and upstream dependency information on top of in-context examples can significantly improve the root cause recommendation accuracy over State-of-The-Art (SoTA) finetuned and in-context learning methods. 

For monitor categorization, our goal is to classify the automated watchdogs or monitors based on resource and SLO classes. Motivated by Srinivas \emph{et al.} \cite{Srinivas2024IntelligentMonitoring}, we considered 13 resource classes and 9 SLO classes. However, in addition to monitor metadata, we curate the service and component properties which are essential to capture the context of a monitor. Our experimental results on 260 real-world monitors demonstrate that additional service and component details in GPT-4 prompt improves the overall accuracy and F1 score both by 4\% over
SoTA methods for SLO classification tasks. To that end, our key contributions are as follows:

\begin{itemize}
    \item This is the first work to demonstrate the promise of using X-lifecycle service information for improving the accuracy of incident management tasks by leveraging real-world dataset from \CompanyX.  
    \item For root cause analysis, we demonstrate that leveraging service functionality and upstream dependency information can assist LLMs in better reasoning and improve the quality of the recommendations over the SoTA methods.
    
    \item For monitor categorization, we demonstrate that leveraging service and component functionalities can boost the classification accuracy while predicting the resource and SLO classes.
    {The results show a consistent enhancement in SLO class predictions while incorporating additional context from service descriptions.}
\end{itemize}
\section{Background} \label{background}
In this section, we provide an overview of incident management lifecycle and highlight the promise of LLMs for automated root cause analysis and monitor categorization. 
\subsection{Incident Management}
\begin{figure}[ht]
    \centering
    \includegraphics[width=0.45\textwidth]{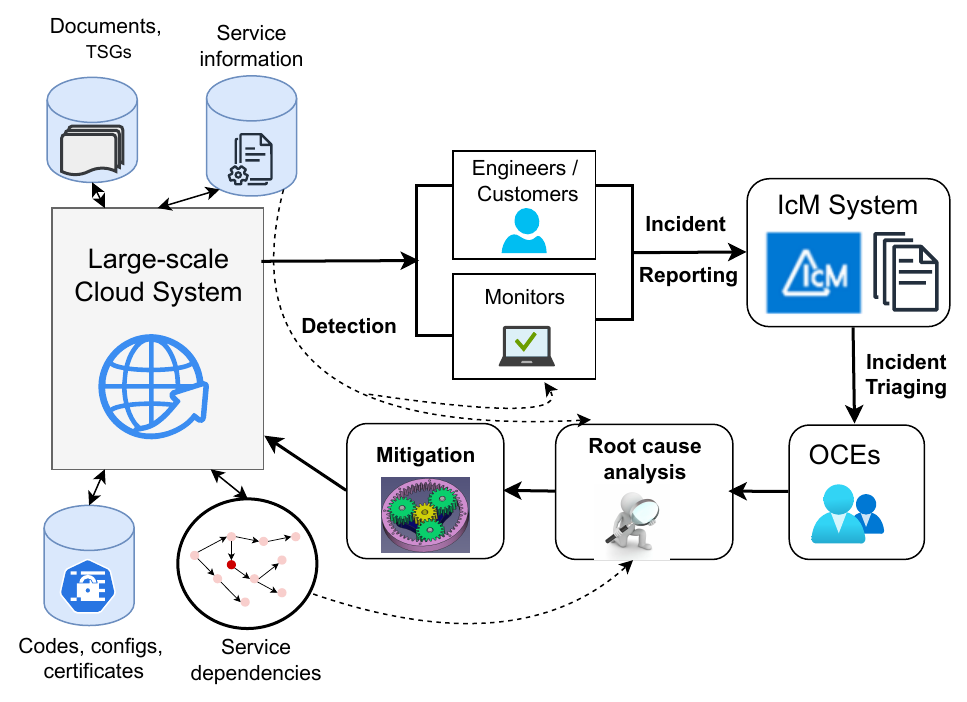}
    \caption{Incident management lifecycle.}
    \Description{Incident management lifecycle.}
    \label{fig:lifecycle}
\end{figure}

Large cloud systems continuously runs tens of thousands of services. Each service stores several important information ranging from service details (\eg, properties, architecture and functionality) to application files (codes, configuration, security certificates), to service dependency data, to troubleshooting guides, and other documentations. Once a service disruption or incident occurs, OCEs carefully refer to these information to properly understand and resolve them. 
Every production incident within \CompanyX{} typically goes through four stages as shown in~\cref{fig:lifecycle}.
\begin{enumerate}
    \item \textbf{Detection:} To quickly detect the failures impacting a service, engineers setup automated watchdogs that continuously monitors the system's health. The incidents reported by these monitors are referred as monitor-reported incidents (MRIs). Internal engineers or external customers of a given service can also report incidents, which are referred as customer-reported incidents (CRIs).
    \item \textbf{Triaging:} Once an incident is reported in a centralized incident management (IcM) system, it is routed to the appropriate team of OCEs after an initial investigation. This incident triaging process sometimes can take multiple rounds to reach the correct team of OCEs.
    \item \textbf{Root cause analysis:} After the incident is triaged to a proper team, multiple rounds of back and forth communication takes place between the OCEs inspecting different aspects of the incident. OCEs usually referred to various source of information including trouble shooting guides and similar historical incidents to understand key root causes of the failure.
    \item \textbf{Mitigation:} Once the root cause is identified, OCEs take several sequential actions resolve the problem and recover the service health.
\end{enumerate}

\subsection{Monitor Management in Cloud Services}
Cloud service owners must continuously monitor their services to maintain high availability and reliability. 
Inadequate monitoring can result in delayed incident detection and negative customer impact ~\cite{Srinivas2024IntelligentMonitoring}. 
Currently, developers rely on a trial and error approach and their tribal knowledge to create monitors, which is ad-hoc and reactive. 
This often leads to incomplete coverage, causing production issues or redundancy, resulting in wasted effort and noise. Therefore, to ensure cloud service reliability, a monitor recommendation framework is crucial, and we can use data-driven approaches for this purpose \cite{Srinivas2024IntelligentMonitoring}. 
However, the unstructured nature of monitors poses a challenge in moving towards a data-driven approach. 
The monitor metadata does not make it obvious which resource characteristics and exact metrics are being monitored (c.f. \cref{fig:sample_monitor}).
Hence, we need a structured approach to tackle this problem, and a structured ontology over the existing monitors is the first step in generating monitor recommendations for a given cloud service. 
By implementing a structured ontology over the monitor data from existing services, we can identify the resource characteristics and exact metrics being monitored for a given service. This structure over the existing monitors and the properties of those services can be leveraged for a data-driven approach to suggest monitors for a new service with a given set of properties.

Srinivas \etal~\cite{Srinivas2024IntelligentMonitoring} presents an empirical study on the existing monitors in \CompanyX{} and defines an ontology over them. The ontology focuses on two dimensions: what to monitor (resource class) and which metrics to monitor (SLO class). The major resource classes include \emph{API}, \emph{Dependency}, \emph{CPU}, \emph{Compute cluster}, \emph{Storage}, \emph{Ram-memory}, \emph{Cache-memory}, \emph{Container}, \emph{Certificates}, \emph{IO}, \emph{Service level}, and \emph{Paging memory}. The major SLO classes include \emph{Success Rate}, \emph{Capacity}, \emph{Latency}, \emph{Availability}, \emph{Throughput}, \emph{Success Rate-QoS}, and \emph{Interruption Rate}. Interested readers can refer to the citation for the definitions of these classes. The current approach for ontology assignment from the citation relies solely on the monitor metadata. However, as seen in the monitor snapshot \cref{fig:sample_monitor}, the monitor metadata contains many software and organization-specific tokens. Additional context on the underlying service might be helpful for better assignment.

\begin{figure}[t]
\ffbox{
\textbf{Monitor Name:} ProcessedDBIngestionValidation  \\
\textbf{Metric Name:} missedSLA \\
\textbf{Service Name:} Service A \\
\textbf{Alert title:} Monitor \{Monitor.DisplayName\} of \\\{Monitor.Dimension.scenario\} is unhealthy \\
\textbf{Alert Conditions:} [\{``Name'':``missedSLAAlert'',``Severity'':4,\\``HealthStatus'':1,``Metadata'':\{\},``Conditions'':[\{\}]\}]
}
\vspace{1mm}
\caption{A sample monitor snapshot.}
\label{fig:sample_monitor}
\end{figure}

\subsection{Root Cause Analysis of Incidents}
Incident root-causing is a complex and challenging process that demands a significant amount of manual effort and domain knowledge about the services involved. Incidents can arise from various failures such as code bugs, configuration errors, dependency failures, hardware failures, and more. Due to the vast number of possibilities, it is non-trivial for OCEs to quickly reach an accurate conclusion. Therefore, root cause analysis is a time-consuming process, and human error can further delay the resolution process, leading to a significant impact on customer satisfaction.

Our primary focus lies on incidents caused by dependency failures from partner services.
\cref{fig:sample_incident} illustrates a real incident stemming from service dependency failures, along with their respective title, summary, and root cause information.
Recently, LLMs have been leveraged to generate automated root cause recommendations at the time of incident creation, aiming to guide the OCEs in the right direction to accelerate the resolution process.

For generating automated root cause recommendations, Ahmed~\etal~\cite{ahmed2023recommending} proposed fine-tuning a LLM model with historical incident data, while Zhang~\etal~\cite{zhang2024automated} suggested leveraging pre-trained LLM models with similar in-context examples.
However, existing research only utilizes incident metadata (\eg, title and summary), which may not always provide accurate information regarding root causes, as shown in~\cref{fig:sample_incident}.
For instance, in the case of dependency failures, it is crucial to carefully inspect all the upstream dependent service properties to understand the critical points of failures.
\begin{figure}[t]
\ffbox{
\textbf{Title:} FuseBot has found a root cause for an anomaly in signal Teams Authentication Service Availability  \\
\textbf{Summary:} The incident has affected the availability and performance health of the Teams Authentication Availability service, as indicated by the Availability Dashboard. Unfortunately, the initial summary does not provide specific error codes or log details, nor does it mention any external services or tools that might contain relevant information. \\
\textbf{Owning Service Name:} Teams Auth Service  \\
\textbf{Root cause:} The root cause of the incident was an Azure networking outage, which led to increased latencies and timeouts with Redis connections in the Teams Auth Service.
}
\vspace{1mm}
\caption{A sample dependency failure incident.}
\label{fig:sample_incident}
\end{figure}
 
\section{Methodology } \label{methods}
In this section, we present our methodology for augmenting LLM prompts with X-lifecycle. contextual information for improving performance of recommendations for two scenarios: (1) Root cause analysis of dependency failure incidents, where service dependencies and their properties play a crucial role; and (2) Monitor classification for resource and SLO classes, where service architecture and functionalities are used to boost the classification accuracy.

\subsection{Root Causing for Dependency Failures}

\subsubsection{Problem Formulation}
Large-scale cloud service providers deal with thousands of incidents daily, originating from different services and teams, of varying severity, which are detected manually or by monitors. Even though the OCEs are highly trained and follow a systematic protocol, finding the root cause of an incident is a non-trivial task.
It not only requires domain-specific knowledge, but also an overall understanding of the intra and inter-service architecture in order to identify how a service failures can trigger a cascading effect and impact other dependent services. Moreover, after having identified a particular causal link, pinpointing the exact cause of failure is a highly complicated process, requiring the input of multiple teams. In the case of such incidents, minimizing the overall impact time is crucial in order to avoid large-scale customer dissatsfaction and uphold the quality of service. 


Due to its practical importance, several research works leverages LLMs for generating automated root causes recommendations either by finetuning an LLM with incident metadata~\cite{ahmed2023recommending} or by querying a pre-trained LLM with similar in-context historical incident examples\cite{zhang2024automated, chen2023empowering}.
However, in practice, OCEs refer to several other contextual information (\eg, logs, codes, service information, monitor reports, dependency information, \etc) while identifying root causes of an incident. 


As there could be several types of root causes, and the contextual data required for proper analysis might differ, we primarily focused on a particular types of incidents which were caused due to dependency failures as it is one of the most frequently encountered causes.
For dependency failures, OCEs typically refer to other upstream services dependencies to identify the critical service owners so as to collaboratively solve the problem.
An \emph{upstream service dependency} denotes a relationship between two services, where one service (the dependent service) relies on the other (the upstream service) to fulfil its operational requirements. A failure, or modification in the upstream service could result in an incident occurring in the dependent service due to the propagating effects.
To accurately identify an incident as an upstream service dependency failure, the OCEs leverage contextual information, their understanding of the affected service, it's dependency architecture, upstream service properties and external factors that could potentially cause the incident, along with the incident metadata, for diagnosis.

In order to emulate the strategy of OCEs, we first retrieve top 5 similar incidents that can be used as in-context examples~\cite{zhang2024automated}. In addition, for a given incident originated from a particular service, we curate all the upstream services and their corresponding service properties, which are then augmented with the in-context examples in the prompt used for querying pre-trained LLM models.

\subsubsection{Data Preparation} 

The data preparation pipeline can be broadly categorized into two parts, (i)~Data collection and (ii)~Data cleaning and summarization. 

(i)~Data collection: We begin by collecting incident data for 353 resolved or mitigated incidents (roughly half of them are caused due to dependency failures) occurred between January 1, 2023 to January 10, 2024 from the IC3 service. Each incident metadata contains it title, initial summary written at the time of incident creation, owning service name and the ground truth root causes.
For a particular incident, we also fetch all the upstream dependency service identifiers (\emph{UpstreamServiceIds}) of its owning service from Microsoft's internal dependency tracking system (DTS) tool. 
This tool curates several information to link the services including shared subscription information, logs of service communication using domain name system(DNS), shared resource information, \etc 
The \emph{UpstreamServiceIds} are then used to fetch the service details from another database that stores the service properties and functionalities.
In addition, we created an retrieval augmented generation (RAG) pipeline where historical incident data are stored in a vector database (as proposed by Zhang \etal~\cite{zhang2024automated}). We then utilize the FAISS \cite{johnson2019billion} library for efficient similarity search and retrieval of top 5 similar in-context incident examples with their corresponding title, summary and reference root causes.

(ii)~Data Cleaning and Summarization: The raw incident summary and root causes that are collected from incident management (IcM) portal contains various noise such as HTML tags, images, tables, stack traces and other non-contextual information, which are not ideal for us due to two reasons: (a)~it can be lengthy and cause input token limit exhaustion during LLM inference, and (b)~additional noise can hamper the reasoning capabilities of pre-trained LLMs and lead to hallucination. Therefore, we used a two-stage method for effectively remove noise from the data. First, we locally process the data to remove HTML tages, tables, long stack traces and image tags. After the initial processing, we summarize the root causes and initial summary in a concise form with the help of GPT-3.5-turbo~\footnote{\href{https://platform.openai.com/docs/models/gpt-3-5-turbo}{https://platform.openai.com/docs/models/gpt-3-5-turbo}} model, where we consider a similar instruction prompt introduced by Zhang \etal~\cite{zhang2024automated}. In addition, we also leverage GPT-3.5-turbo model to summarize a subset of service descriptions which were verbose and noisy. A sample instruction prompt for service property summarization is illustrated in \cref{fig:ServiceDescription_summarizationPrompt}.



\begin{figure}[t]
\ffbox{
I want you to act as an expert software engineer. Consider the service descriptions. 

\vspace{1mm}
\textbf{Upstream service 1 -} Description

$\vdots$

\textbf{Upstream service n -} Description

\vspace{1mm}
Your task is to summarize these service descriptions.\\ Focus on the following aspects of the service functionality: \\
* The functionality provided by the service \\
* References to resources it operates \\
* Distinguishing features of the service that clearly explains its operation\\
Your summary should be at most 2-3 sentences and should be in third person.
}
\vspace{1mm}
\caption{Prompt to summarize the Service Descriptions}
\label{fig:ServiceDescription_summarizationPrompt}
\end{figure}

\subsubsection{Methods and Baselines}\label{subsubsec:methods}

To critically observe the impact of upstream service information in the root-cause analysis, we develop five strategies for leveraging LLM models~\footnote{Except for the finetuned model, we always perform inference with GPT4 model, where the \emph{temperature} parameter is set to 0 in order to get deterministic results.}:  

\begin{figure}[htbp!]
\ffbox{
\textbf{-- Task Description:}

\vspace{1mm}
- You are dealing with root cause analysis for Azure cloud incident. You are an \textbf{**on-call engineer**} that is responsible for investigating the root cause.

- You will be provided with title and summary of the incident along with the service name and its functionality that affected by the incident.  

- You will also be provided with a list of \textbf{** Upstream service dependencies**} and their \textbf{**Descriptions**}

- In the context of software engineering and system architecture, an upstream service dependency denotes a relationship between two services, where one service (the dependent service) relies on another service (the dependency) to fulfill its operational requirements. This dependency typically involves the exchange of data, execution of functions, or access to resources necessary for the dependent service to perform its intended tasks. This relationship is termed ``upstream'' to signify the flow of dependencies from the dependent service to its sources, akin to the flow of a river from its upstream source. 

- The Upstream service dependencies have been provided so that you can better identify the root cause of the incident.

\vspace{1mm}
\textbf{-- Historical Incident Examples:}

\vspace{0.5mm}
- Below are some historical incidents with their root causes. If you find the description and title to be \textbf{**similar or relevant**} to the current incident, you can also use the examples to determine the root cause of the current incident.

\vspace{0.5mm}
- Historical Incident Summary 1: {Retrieved summary}

- Historical Incident Title 1: {Retrieved title}

- Historical Incident Root Cause 1: {Retrieved Root cause} \\
$\vdots$

- Historical Incident Summary 5: {Retrieved summary }

- Historical Incident Title 5: {Retrieved title}

- Historical Incident Root Cause 5: {Retrieved Root cause}

\vspace{1mm}
\textbf{-- Answering Format:} Your output response should strictly be a **Json** file with the following two objectives:

\vspace{0.5mm}
\textbf{Objective1:} Infer and explain in detail, the Root cause that could have caused the incident 

\textbf{Objective2:} Binary classification as follows: If and only if you identify the root cause as an upstream service dependency, return ``Yes''. Else, if the incident is \textbf{**not**} caused due to the failure of an upstream service, return ``No''

\vspace{1mm}
\textbf{-- Incident Details:}  \\
- Service: {Service Name} 

- Functionality: {Functionality of the Service} 

- Incident Summary: {The summary of the incident} 

- Incident Title: {Title of the incident} 

\vspace{1mm}
\textbf{-- Upstream Service Dependencies:}

\vspace{0.5mm}
1. Upstream Service Name – {Description of the Upstream Service}

$\vdots$ 

n. Upstream Service Name – {Description of the Upstream Service}
}
\vspace{1.5mm}
\caption{Prompt with Incident Information, In-context Examples, and Upstream Service Dependency Details (InC DEP)}
\label{fig:inc_w_dep}
\end{figure}

\begin{enumerate}
    \item \textbf{Our method (InC DEP) -- Incident Info + In-context Examples +  Dependency Details} develops a prompt with five distinct parts: \emph{Task description}, \emph{In-context examples}, \emph{Answering format}, \emph{Incident details} and \emph{Upstream service information}. An illustration of the prompt is demonstrated in \cref{fig:inc_w_dep}. The \emph{Task Description} instructs the model to assume the role of an OCE to perform root cause analysis for Azure cloud incidents, followed by describing the concept of \emph{Upstream service dependencies and failures}. For in-context examples, we retrieve the top 5 similar incidents from historical data corpus using  FAISS \cite{johnson2019billion} library, and add their corresponding title, summary and root causes.
    
    The answering format specifies the format in which the model should give the output. We instruct the model to return the answers in a \emph{json} format with two \emph{Objectives}: (a)~inferring the potential root cause of the incident and (b)~based on recommended root cause, classify whether it is a dependency failure. The incident details provides contextual information such as the title, summarized initial summary, owning service name and a summary of its functionality.
    
    Lastly, the upstream service information contains the list of upstream service dependencies of the owning service and their corresponding summarized service properties. Thus, we query a pre-tained GPT-4 model to reason with incident metadata, in-context examples and upstream service properties, to infer the potential root causes of a given incident.
    
    \item \textbf{Only Incident Info (NoDEP).} For this baseline method, we develop a prompt only with incident metadata (\ie title, summary and owning service description) and query the GPT-4 model in a zero-shot setting.

    \item \textbf{Incident Info + Dependency Details (DEP).} In this prompting strategy, in addition to the incident metadata, a list of all the \emph{Upstream Service Dependencies} and their summarized descriptions are also included. The \emph{Task Description} section also includes statements that introduce the \emph{Upstream Service Dependencies} section and it's objective.
    
    \item \textbf{Incident Info + In-context Examples (InC NoDEP)~\cite{zhang2024automated}.} For this method, we emulate the prompt strategy proposed by Zhang \etal~\cite{zhang2024automated}, where along with incident metadata, we also retrieve the top 5 similar in-context incident examples and add their corresponding metadata into prompt. In addition, we provide a summarized description of owning service properties in the prompt.

    \item \textbf{Finetuning LLM~\cite{ahmed2023recommending} (FtGPT).} Lastly, We finetune a GPT-3 model with last 3 years of historical incidents originated from all the services within \CompanyX, where the title, summary and owning tenant name are given as input and the ground truth root causes are provided as reference. The fine-tuned GPT-3 model is expected to learn the semantics of the incident management domain by feeding information regarding all possible types of information and their root causes.

\end{enumerate}

\vspace{-1.5mm}
\subsubsection{Evaluation metrics} We use both \emph{lexical} and \emph{semantic} quantitative metrics to evaluate similarity between actual root cause and the recommendations generated by different models.
 
\paragraph{\textbf{Lexical Metrics.}} We use \emph{three} lexical metrics, namely, the smooth sentence BLEU-4 (Bilingual Evaluation Understudy) \cite{lin-och-2004-orange-bleu} metric to assess the overlap of n-grams ranging from 1 to 4 between the reference and generated texts. Also, we apply the ROUGE metric (Recall Oriented Understudy for Gisting Evaluation) \cite{lin-2004-rouge} to compare a candidate document with a set of reference texts. We specifically pick ROUGE-L \cite{lin-2004-rouge}, which accounts for sentence-level structural similarity and finds longest co-occurring in sequence n-grams based on Longest Common Subsequence (LCS) \cite{lcs} statistics. Furthermore, we incorporate the METEOR (Metric for Evaluation of Translation with Explicit Ordering) \cite{banerjee-lavie-2005-meteor} which evaluates the generation quality based on the harmonic mean of unigram precision and recall as well as stemming and synonymy matching as extra features.

\paragraph{\textbf{Semantic Metrics.}} Unlike lexical metrics, which primarily focus on exact word matching, we want to evaluate the preservation of overall semantic meaning in the generated results. To achieve this, we use \emph{two} semantic metrics that go beyond exact word matches and consider the contextual meaning of words. We use BERTScore \cite{bert-score}, which uses the pre-trained contextual embeddings from the BERT \cite{devlin-etal-2019-bert} model to measure the similarity between candidate and reference sentences based on cosine similarity. 
Lastly, we integrate NUBIA (NeUral Based Interchangeability Assessor) \cite{kane-etal-2020-nubia}, a recent neural-based measure that combines the semantic similarity, logical inference, and sentence legibility from exposing layers of pre-trained language models, including RoBERTa STS \cite{Liu2019RoBERTaAR}, RoBERTa MNLI and GPT-2 \cite{Radford2019LanguageMA-GPT2}.

\subsection{Monitor Management}
\subsubsection{Problem Formulation}
{Detecting anomalous behavior or early symptoms can significantly reduce the impact on customers and speed up the process of resolving incidents.
At \CompanyX, production incidents are either detected by automated system watchdogs that continuously monitor service health and telemetry data, or reported by users (internal or external).
Although literature highlights the importance of monitoring Service Level Objectives, it is often not clear what exactly to monitor and where to monitor for a given service \cite{Srinivas2024IntelligentMonitoring}.
Service owners or developers typically use their expertise and domain knowledge to create monitors that ensure the continuous availability of a service.
Given that each monitor name differs even if it observes the same signal, making it essential to identify and learn the underlying structure of monitors to move towards an automated approach to monitor recommendation using service properties.}
{
In their work, \cite{Srinivas2024IntelligentMonitoring} defines two major dimensions of the monitors: resource classes (what to monitor within a service) and SLO classes (which metrics are most representative of the performance, efficiency, and reliability of a cloud service) to restructure the monitor space.
As fine-tuning is resource-intensive and requires significant human effort in generating labels, \cite{Srinivas2024IntelligentMonitoring} learns structure over the space using in-context learning with monitor metadata.

However, learning structure over the monitor space given the monitor metadata is not a trivial task.
In addition to the domain-specific knowledge, this task also requires additional context such as the exact nature of the underlying service.
Therefore, our aim here is to provide additional context by combining data that exists in silos  to structure the monitor space with better accuracy.
Our problem here boils down to given the monitor metadata, leverage additional context from various silo databases to identify the resource and SLO classes associated with the monitor.
}
\subsubsection{Data Preparation}
{
The company maintains an internal directory that tracks all services, providing information on the functionality description of each service, as well as the microservices and components within them. Therefore, we curate the following information for each monitor:
\begin{enumerate}
\item Contextual information on the exact functionality of the underlying service
\item Contextual information on the service architecture, namely, service components and their individual functionality.
\end{enumerate}
}
{We combine the monitor metadata, which includes
monitor name, metric name, service name, alert title (used to create the alert when the thresholds are violated), and the alert conditions, along with the service description and descriptions of components associated with the service. We leverage the contextual information to mine for the monitor structure.}
{
We manually labeled 260 monitors for resource classes and 180 monitors for SLO Classes. This labeled data served as the ground truth for our predictive models.\\
}
\begin{figure}[t]
\ffbox{
\textbf{-- Task Description:}

\vspace{1mm}
\noindent - You are an intelligent virtual assistant that answers questions from a user based on the monitor metadata provided. Go through monitor metadata \{'Monitor Name', 'Account Name', 'namespace', 'Metric name', 'Sampling Type', 'Alert Title', 'Service Name'\} , preprocess the text such as camel casing, snake case splitting, etc. and predict the resource class of the monitor. Answer the following questions about the monitor sequentially.

\vspace{0.5mm}
1. Q1: Based on the monitor description, identify the underlying entity being tracked. Pay attention to features, internal functions, dependencies, datastores, or service level objectives. 

\vspace{0.5mm}
2. Q2: Categorize the identified entity into a generic class:
\vspace{0.5mm}
\begin{enumerate}
    \item api
    \item dependency
    \item compute cluster 
    \item service level 
    \item cache-memory 
    \item ram-memory 
    \item CPU 
    \item paging memory 
    \item container 
    \item IO 
    \item Storage 
    \item none-of-the-above 
\end{enumerate}

\vspace{0.5mm}
\textbf{-- Additional Guidance :} \\
- Focus on extracting relevant information from the descriptions.

- Pay attention to specific features or characteristics indicative of the resource class. 

- Consider negations or exceptions in descriptions. 

- Highlight the importance of understanding relationships between different elements. 
}
\vspace{1.5mm}
\caption{Prompt for generating resource class labels}
\label{fig:resource_class}
\vspace{-0.5mm}
\end{figure}

\vspace{-3mm}
\subsubsection{Methods and Baselines}
{We use GPT4 model and conduct experiments with various combination of data sources to evaluate their efficacy in generating resources and SLO classes:}
{
\begin{enumerate}
    \item Case 1: Baseline experiment only utilizes monitor metadata for prediction.
    \item Case 2: Incorporating monitor data and service descriptions.
    \item Case 3: Combining monitor data, service descriptions, and descriptions of components associated with the service.
    \item Case 4: Employing monitor data and descriptions of components associated with the service.
\end{enumerate}
}

 \subsubsection{Evaluation metrics}
{
We assess the performance of contextual information on each of the resource and SLO classes using precision, recall, F1-score, and accuracy metrics. 
Metrics were calculated for each class within resource and SLO class categories. 
By evaluating these metrics across different cases, we aim to understand the impact of incorporating additional data on the accuracy and effectiveness of predicting resource and SLO classes. 
}
\section{Experimental Results} \label{results}



In this section, we present the performance of our proposed methodology for two earlier mentioned scenarios: (a) How dependent-service description help in improving root cause analysis? and (b) How service architecture and functionality information help in monitor categorization? 
\subsection{How dependent-service description help in improving root cause analysis?}

For this experiment, we used a data corpus with 3 years of incidents and selected 353 recent incidents from IC3 for evaluation, while rest are used for creating vector database for in-context learning and finetuning the GPT model. In terms of classifying whether an incident is caused due to dependency failures, we observe that without dependency information (i.e., NoDEP method), we achieve a F1-score of 0.4, while adding in-context examples (i.e., InC NoDEP method) add more noise and further reduces the F1-score to 0.16. However, adding dependency information in the prompt (i.e., DEP and InC DEP method) can boost the F1-score to 0.58 while classifying dependency failures.

In \cref{tab:rca}, we demonstrate the effectiveness of our model in employing both \emph{upstream service description} as well as \emph{in-context} description. We conducted a comparison across four prompting strategies along with FinetunedGPT \cite{ahmed2023recommending}, as outlined in \cref{subsubsec:methods}. From these results, we identify a few interesting observations:
(1) Expect for NUBIA scores, the average similarity metrics for finetuned GPT-3 model recommendations are almost always lower than other methods. This can be due to two factors: (a) we finetuned a GPT-3 model as finetuning a GPT-4 model requires significant cost and GPU resources, whereas other methods leverages GPT-4 model for inferencing, and (b) as finetuning considers incidents from all the services, it is prone to hallucinate for these specific dependency failures. 
(2) Providing only incident description (i.e., NoDEP method) does not help in imparting right contextual information regarding the consequences resulting from its failure, and therefore, performs poorly. Having service description along with incident metadata in prompt (i.e., DEP method) also does not help in improving the quality of root-cause recommendations, rather the performance drops slightly in some cases.
(3) Once we include in-context examples in the prompt without dependencies (i.e., InC NoDEP), usually it sightly outperforms DEP, NoDEP and FtGPT method, as similar examples help the model to better reason about current incident and reason about how to write the root cause recommendation.
However, once we add the service dependency information along with in-context examples, we observe that the model now has full context about what upstream dependent service might cause the failure along with how to frame the root causes for such failures. Therefore, our method (InC DEP) boost the performance by a huge margin in comparison to other benchmark methods. For lexical metrics, we observe a 5-38\% gain (lowest for METEOR; highest for BLEU), while for the semantic metrics NUBIA scores gains by 54.67\% over other promting strategies.
\begin{table}[t]
\centering
\tabcolsep=0.05cm
\small
\begin{tabular}{l ccccc}
\toprule
           & FtGPT~\cite{ahmed2023recommending}    & NoDEP   & DEP     & InC NoDEP & InC DEP \\ \midrule
BLEU       & 5.64 \scriptsize{$\pm$ 5.05} & 6.26 \scriptsize{$\pm$ 2.14}  & 6.14 \scriptsize{$\pm$ 2.55}  & 6.86 \scriptsize{$\pm$ 2.69} & \textbf{9.43} \scriptsize{$\pm$ 4.23}  \\
METEOR     & 20.14 \scriptsize{$\pm$ 8.27} & 26.41 \scriptsize{$\pm$ 6.16} & 26.03 \scriptsize{$\pm$ 6.58} & 26.54 \scriptsize{$\pm$ 6.98}   & \textbf{26.72} \scriptsize{$\pm$ 8.56} \\
ROUGE      & 17.72 \scriptsize{$\pm$ 9.75} & 16.32 \scriptsize{$\pm$ 3.66} & 16.03 \scriptsize{$\pm$ 4.11} & 17.59 \scriptsize{$\pm$ 4.35}   & \textbf{21.14} \scriptsize{$\pm$ 6.34} \\
BERT Score & 85.60 \scriptsize{$\pm$ 3.04} & 85.83 \scriptsize{$\pm$ 1.40} & 85.73 \scriptsize{$\pm$ 1.48} & 86.11 \scriptsize{$\pm$ 1.47}   & \textbf{86.65} \scriptsize{$\pm$ 1.67} \\
NUBIA      & \textbf{20.63} \scriptsize{$\pm$ 20.12} & 12.32 \scriptsize{$\pm$ 17.26} & 11.68 \scriptsize{$\pm$ 15.12} & 13.17 \scriptsize{$\pm$ 18.14}   & \emph{20.37} \scriptsize{$\pm$ 21.70} \\ \bottomrule
\end{tabular}
\caption{Performance of 4 different prompting strategies along with the finetuned model on 5 different quantitative metrics with mean and standard deviations. \emph{DEP}: denotes the case where we provide descriptions for upstream-service-dependencies, while in \emph{NoDEP} we don't provide any. \emph{InC}: refers to the case where we have In-Context examples as well in the prompt. \emph{FtGPT}: denotes FinetunedGPT.}
\label{tab:rca}
\vspace{2mm}
\end{table}

Further, to evaluate the overall performance regarding incidents stemming from upstream service failures versus other causes, we segregated the incidents into two distinct categories based on ground truth of incident categories available in IcM.
As depicted in Table \ref{tab:rca_comp}, we demonstrate the performance comparison for different quantitative metrics for incidents caused due to service dependency failures (referred to as SD) vs other categories of incidents (referred to as NoSD). As expected, we observe that the performance of our InC DEP method is significantly better for generating close to ground truth root cause recommendations for dependency failure incidents over other categories of incidents.
Moreover, except for METEOR score, we observe that the performance gain generalizes to other categories of incidents as well, as our InC DEP method almost always outperforms other prompting strategies for non-dependency related failures as well.

\begin{table*}[t]
\centering
\tabcolsep=0.32cm
\begin{tabular}{l cccccccccc}
\toprule
\multirow{2}{*}{} &
  \multicolumn{2}{c}{FtGPT~\cite{ahmed2023recommending}} &
  \multicolumn{2}{c}{NoDEP} &
  \multicolumn{2}{c}{DEP} &
  \multicolumn{2}{c}{InC NoDEP} &
  \multicolumn{2}{c}{InC DEP} \\ \cmidrule(lr){2-3} \cmidrule(lr){4-5} \cmidrule(lr){6-7} \cmidrule(lr){8-9} \cmidrule(lr){10-11}
 &
  \multicolumn{1}{c}{NoSD} &
  \multicolumn{1}{c}{SD} &
  \multicolumn{1}{c}{NoSD} &
  \multicolumn{1}{c}{SD} &
  \multicolumn{1}{c}{NoSD} &
  \multicolumn{1}{c}{SD} &
  \multicolumn{1}{c}{NoSD} &
  \multicolumn{1}{c}{SD} &
  \multicolumn{1}{c}{NoSD} &
  \multicolumn{1}{c}{SD} \\ \midrule
BLEU       & 5.271           & 5.972  & 6.268  & 6.255  & 6.085  & 6.190  & 6.884  & 6.843  & \textit{9.285}  & \textbf{9.551}  \\
METEOR     & 19.448          & 20.750   & 26.048 & 26.731 & 25.389 & 26.584 & 25.921 & \emph{27.087} & 25.708          & \textbf{27.605} \\
ROUGE      & 17.364          & 18.035 & 16.416 & 16.234 & 16.045 & 16.010 & 17.502 & 17.670 & \textit{20.644} & \textbf{21.577} \\
BERT Score & 85.671          & 85.523 & 85.790 & 85.865 & 85.578 & 85.858 & 86.017 & 86.199 & \textit{86.508} & \textbf{86.772} \\
NUBIA      & \textbf{23.050} & 18.511 & 13.578 & 11.210 & 12.733 & 10.762 & 15.141 & 11.437 & \textit{22.982} & 18.072 \\ \bottomrule
\end{tabular}
\caption{Performance comparison for different methods for incidents caused due to an upstream dependency failure (SD) vs. incidents caused due to non-dependency failures (NoSD).}
\label{tab:rca_comp}
\vspace{1mm}
\end{table*}

\paragraph{\textbf{Human evaluation.}} 
We also manually evaluated the accuracy of root cause recommendations from different methods against the ground truth root cause information. For this study, we chose 30 recent incidents from our pool of 353 incidents, out of which 20 were caused due to dependency failures. This evaluation involves three of the authors first separately providing correctness score for each of the recommendations in the scale of 1 to 5, where "5" represents that the recommendation fully captures the semantic of ground truth and "1" represents that the recommendation is completely irrelevant. Finally, all three evaluators discuss and finalize the scores where there were disagreements.
The overall correctness scores from recommendations of different methods are summarized in~\cref{tab:human_eval}.
Overall, we observe that our method with dependency information and in-context examples significantly outperforming the FinetunedGPT model by 53.89\%. Moreover, additional dependency information can further boost the accuracy by 3.6\% over the InC NoDEP method.
\begin{table}[t]
\centering
\tabcolsep=0.12cm
\begin{tabular}{l ccccc}
\toprule
                & FtGPT      & NoDEP       & DEP         & InC NoDEP   & InC DEP     \\ \midrule
$\mu \pm \sigma$ & 1.8 \scriptsize{$\pm$ 1.1} & 2.27 \scriptsize{$\pm$ 1.17} & 2.63 \scriptsize{$\pm$ 1.13} & 2.67 \scriptsize{$\pm$ 1.24} & \textbf{2.77} \scriptsize{$\pm$ 1.36} \\
Median          & 1         & 2           & 2           & 2           & 2           \\
Mode            & 1         & 1           & 2           & 2           & 2           \\ \bottomrule
\end{tabular}
\caption{Correctness score (Mean ($\mu$) $\pm$ Standard-deviation ($\sigma$), Median, and Mode) from human evaluation.}
\label{tab:human_eval}
\vspace{2mm}
\end{table}

\subsection{How service architecture and functionality information help in monitor categorization?}
Here, we analyze the effect of different contextual information on identifying monitor classes and how this effect varies across resource and monitor classes.

\vspace{-3mm}
\subsubsection{{How can service architecture and functionality help improve the accuracy of identifying the SLO categories of monitors?}}
\begin{table*}[t]
\begin{tabular}{l cccccccccccccccc}
\toprule
            & \multicolumn{4}{c}{precision}                                              & \multicolumn{4}{c}{recall}                                                 & \multicolumn{4}{c}{f1-score}                                               & \multicolumn{4}{c}{accuracy} \\\cmidrule(lr){2-5}\cmidrule(lr){6-9} \cmidrule(lr){10-13}\cmidrule(lr){14-17} 
            & C1 & C2 & C3 & C4 & C1 & C2 & C3 & C4 & C1 & C2 & C3 & C4 & C1 & C2 & C3 & C4 \\\midrule
Availability      & 0.48     & 0.48                & 0.47                           & 0.45                   & 0.85     & \textbf{1.00}                & 0.88                           & 0.88                   & 0.59     & 0.65                & 0.61                           & 0.60                   & 0.85     & \textbf{1.00}                & 0.88                           & 0.88                   \\
Capacity          & 0.89     & 0.90                & 0.90                           & 0.89                   & 0.57     & 0.64                & 0.64                           & 0.57                   & 0.70     & 0.75                & 0.75                           & 0.70                   & 0.57     & 0.64                & 0.64                           & 0.57                   \\
Freshness         & 1.00     & 1.00                & 0.90                           & 1.00                   & 0.67     & 0.75                & 0.75                           & 0.67                   & 0.80     & 0.86                & 0.82                           & 0.80                   & 0.67     & 0.75                & 0.75                           & 0.67                   \\
Interruption Rate & 1.00     & 1.00                & 1.00                           & 0.50                   & 0.23     & 0.31                & 0.15                           & 0.08                   & 0.38     & 0.47                & 0.27                           & 0.13                   & 0.23     & 0.31                & 0.15                           & 0.07                   \\
Latency           & 0.82     & 0.88                & 0.82                           & 0.73                   & 1.00     & 1.00                & 1.00                           & 0.83                   & 0.90     & 0.94                & 0.90                           & 0.78                   & 1.00        & 1.00                & 1.00                           & 0.82                   \\
Others            & 0.00     & 0.00                & 0.50                           & 0.00                   & 0.00     & 0.00                & 0.14                           & 0.00                   & 0.00     & 0.00                & \textbf{0.22}                           & 0.00                   & 0.00     & 0.00                & 0.14                           & 0.00                   \\
Reliability       & 0.87     & \textbf{0.96}                & 0.88                           & 0.90                   & \textbf{0.94}     & 0.92                & 0.90                           & 0.92                   & 0.90     & 0.94                & 0.89                           & 0.91                   & \textbf{0.94}     & 0.92                & 0.90                           & 0.92                   \\
Success Rate      & 0.85     & \textbf{0.92}                & 0.69                           & 0.73                   & 0.55     & 0.55                & 0.45                           & 0.55                   & 0.67     & 0.69                & 0.55                           & 0.63                   & 0.55     & 0.55                & 0.45                           & 0.55                   \\
Throughput        & 0.87     & 0.87                & 0.86                           & 0.81                   & 0.93     & 0.93                & 0.86                           & 0.93                   & 0.90     & 0.90                & 0.86                           & 0.87                   & 0.93     & 0.93                & 0.86                           & 0.93                  
\\ \bottomrule
\end{tabular}
\caption{ Quantitative metrics evaluated on SLO recommendations }
\label{tab:slo_qual_rec}
\vspace{1mm}
\end{table*}

{\cref{tab:slo_qual_rec} presents class-wise metrics for individual SLO classes with different contextual information. C1 indicates the baseline results, C2 indicates predictions with service descriptions and monitor metadata, C3 indicates predictions with service descriptions, monitor metadata, and component level descriptions, and C4 presents the predictions with monitor metadata and component level descriptions.
}
{For SLO classification, the addition of service description appears to have a particularly positive effect on classes such as \emph{Availability} and \emph{Freshness}, where the recall significantly increased. This suggests that the additional information provided by the service description aids in better identifying instances belonging to these classes.}
 \begin{figure}[htbp!]
\ffbox{
\textbf{--Output 1}

\vspace{1mm}
\textbf{Q1}: The Service Level Objective (SLO) tracked by this monitor appears to be the freshness of data in seconds. This is inferred from the 'Metric name' field which mentions $\{ \}$. This suggests that the monitor is tracking how fresh or up-to-date the data is, measured in seconds. The lower the value, the fresher the data is considered to be. 

\vspace{0.5mm}
The service description also provides some context. It mentions that the service is part of the $\{ \} $ team and they own $\{ \}$. This implies that the freshness of data could be critical for their operations as they need to ensure that the ads being served are based on the most recent and relevant data.

\vspace{0.5mm}
\textbf{Q2}: The SLO being tracked by this monitor can be categorized under the generic class of '7. Freshness'. This is because it is monitoring the 'DataFreshnessInSeconds', which directly relates to how current or fresh the data is.

\vspace{1mm}
\textbf{--Output 2} \\
\textbf{Q1}: The Service Level Objective (SLO) tracked by this monitor appears to be related to the freshness of definitions. The monitor name $\{ \}$ suggests that it is tracking the age of certain definitions, likely related to the services provided $\{ \}$. If these definitions exceed 7 days old, an alert is triggered. This could be important in ensuring that the most up-to-date definitions are being used in the service delivery for $\{ \}$.

\vspace{0.5mm}
\textbf{Q2}: The SLO being tracked by this monitor can be categorized under the generic class of ``Freshness''. This is because the monitor is tracking the age of definitions, which is a measure of how current or fresh the definitions are.
}
\vspace{1.5mm}
\caption{Sample outputs from SLO classification with additional context from service description.}
\label{fig:slo_output}
\vspace{-1mm}
\end{figure}
{\cref{fig:slo_output} shows the GPT response for prompts with additional in context information. As seen in the figure, the additional service description provides better context for SLO classification. }
{
On the other hand, in classes like \emph{Interruption Rate} and \emph{Success Rate}, the impact seems less pronounced, indicating that the service description might not provide as much discriminatory information for these classes. Overall, the results indicate that incorporating service description alongside monitor metadata leads to improved prediction performance for certain SLO classes, particularly in terms of recall and F1-score. 

We observe that the experimental model shows a slight improvement in accuracy compared to the baseline model, with an increase from 0.75 to 0.79. This indicates that incorporating service description contributes to better overall prediction performance across all SLO Classes. 
In terms of the macro average of precision, recall, and f1-score, the experimental model also exhibits improvement, albeit modest. The macro average precision, recall, and f1-score increased from 0.78, 0.75, and 0.73 in the baseline model to 0.83, 0.79, and 0.77 respectively, in the experimental model. This indicates that the experimental model performs better, particularly for the classes that are more prevalent in the dataset.
{While the inclusion of both service and component descriptions in C3 does yield improvements in certain cases such as \emph{Others}, it does not consistently show remarkable enhancements compared to the baseline. This observation suggests that while the overall functionality of a service can indeed define its service level objectives, individual component descriptions, as utilized in C4, may not always contribute significantly to the predictive performance.}

\noindent Overall, while the improvements may vary across different evaluation metrics, the experimental model consistently outperforms the baseline model, demonstrating the effectiveness of integrating additional features for enhanced prediction accuracy and reliability. 
}
\subsubsection{{How can service architecture and functionality help improve the accuracy of identifying the resource categories of monitors?}}
\begin{table*}[]
\begin{tabular}{l cccccccccccccccc}
\toprule
            & \multicolumn{4}{c}{precision}                                              & \multicolumn{4}{c}{recall}                                                 & \multicolumn{4}{c}{f1-score}                                               & \multicolumn{4}{c}{accuracy} \\\cmidrule(lr){2-5}\cmidrule(lr){6-9} \cmidrule(lr){10-13}\cmidrule(lr){14-17} 
            & C1 & C2 & C3 & C4 & C1 & C2 & C3 & C4 & C1 & C2 & C3 & C4 & C1 & C2 & C3 & C4 \\\midrule
API              & 0.44                          & 0.43                & 0.32                           & 0.35                   & 0.72     & 0.83                & \textbf{0.91}                           & 0.87                   & 0.55     & 0.56                & 0.47                           & 0.50                   & 0.78     & 0.83                & \textbf{0.91}                           & 0.88                   \\
Dependency        & 0.64                          & \textbf{0.76}                & 0.67                           & 0.71                   & 0.47     & 0.50                & 0.37                           & 0.51                   & 0.55     & 0.60                & 0.48                           & 0.59                   & 0.50     & 0.54                & 0.38                           & 0.51                   \\
Compute cluster   & 0.69                          & 0.71                & 0.50                           & \textbf{0.89}                   & 0.38     & 0.50                & 0.41                           & 0.43                   & 0.49     & 0.59                & 0.45                           & 0.58                   & 0.47     & 0.52                & 0.42                           & 0.46                   \\
Service level     & 0.48                          & 0.47                & 0.50                           & 0.52                   & 0.33     & 0.22                & 0.20                           & 0.24                   & 0.39     & 0.30                & 0.28                           & 0.33                   & 0.39     & 0.26                & 0.20                           & 0.25                   \\
Cache-memory      & 0.81                          & \textbf{0.90}                & 0.79                           & 0.76                   & 0.85     & 0.95                & 0.96                           & 0.96                   & 0.83     & \textbf{0.93}                & 0.86                           & 0.85                   & 1.00     & 1.00                & 0.96                           & 1.00                   \\
Ram-memory        & 1.00                          & 1.00                & 1.00                           & 1.00                   & 0.83     & \textbf{1.00}                & 0.99                           & 0.99                   & 0.91     & 1.00                & 0.99                           & 1.00                   & 0.95     & 1.00                & 0.98                           & 1.00                   \\
CPU              & 0.63                          & 0.62                & \textbf{0.85}                           & 0.84                   & 0.90     & 1.00                & 0.96                           & 0.95                   & 0.75     & 0.76                & \textbf{0.90}                           & 0.89                   & 1.00     & 1.00                & 0.96                           & 1.00                   \\
Paging memory     & 0.93                          & 0.93                & 0.94                           & 0.94                   & 0.88     & 0.88                & 0.56                           & 0.63                   & 0.90     & 0.90                & 0.70                           & 0.76                   & 0.93     & 0.93                & 0.56                           & 0.63                   \\
Container         & 1.00                          & 1.00                & 1.00                           & 1.00                   & 0.63     & 0.57                & 0.64                           & 0.62                   & 0.77     & 0.73                & 0.78                           & 0.76                   & 0.67     & 0.59                & 0.64                           & 0.63                   \\
IO               & 0.76                          & 0.70                & 0.82                           & 0.82                   & 0.89     & 0.89                & 0.92                           & 0.96                   & 0.82     & 0.78                & 0.87                           & 0.89                   & 0.94     & 0.89                & 0.92                           & 0.98                   \\
Storage           & 0.72                          & 0.59                & 0.64                           & 0.66                   & 0.92     & 0.76                & 0.96                           & 0.96                   & 0.81     & 0.67                & 0.77                           & 0.78                   & 0.92     & 0.83                & 0.96                           & 0.96                   \\
Certificate       & 1.00                          & 0.96                & 1.00                           & 1.00                   & 1.00     & 1.00                & 0.97                           & 1.00                   & 1.00     & 0.98                & 0.98                           & 1.00                   & 1.00     & 1.00                & 0.97                           & 1.00                   \\
None-of-the-above & 0.00                          & 0.33                & 0.75                           & \textbf{0.94}                   & 0.00     & 0.05                & 0.04                           & \textbf{0.22}                   & 0.00     & 0.09                & 0.07                           & \textbf{0.35}                   & 0.00     & 0.06                & 0.04                           & \textbf{0.21}                  

\\ \bottomrule
\end{tabular}
\caption{Quantitative metrics evaluated on Resource Class recommendations }
\label{Tab:Resource class prediction}
\vspace{2mm}
\end{table*}

{\cref{{Tab:Resource class prediction}} presents class-wise metrics for individual resource classes with different contextual information. C1 indicates the baseline results, C2 indicates predictions with service descriptions and monitor metadata, C3 indicates predictions with service descriptions, monitor metadata, and component level descriptions, and C4 presents the predictions with monitor metadata and component level descriptions.
}
{
The incorporation of service description has varying impacts across different resource classes, demonstrating the need for a nuanced approach in improving predictive models. While some classes show substantial improvements like \emph{Dependency} and \emph{API}, others such as \emph{Storage} and \emph{Service level} exhibit limited changes with additional information.

If we look at overall impact on micro, macro and weighted averages, we observe that inclusion of service description in the experimental model is modest. While there are slight improvements in macro-average metrics, there's also a trade-off in micro-average recall. The weighted-average metrics show a marginal increase in precision and recall. 

Incorporating service descriptions has the potential to enhance the interpretability of the model by providing detailed context and enabling more informed decision-making. 
Striking a balance between leveraging the valuable information from service description and maintaining a manageable level of interpretability is crucial.

We also investigated the effect of including descriptions of components associated with the service, in addition to service descriptions. Our analysis revealed that while the inclusion of component descriptions introduces additional information, it does not lead to substantial improvements in predictive performance compared to model utilizing only service descriptions. For many resource classes, the precision, recall, and f1-score remained relatively stable between the two models. This suggests that the additional information provided by component descriptions did not significantly influence the model's ability to accurately classify resource classes. In some cases, we observed slight fluctuations in performance metrics between the two models. However, these changes were not substantial enough to indicate a clear advantage in predictive performance for the model incorporating component descriptions. 

The monitoring setup involves aggregating monitors at the service level rather than at the component level. As a result, the additional granularity provided by component descriptions may not align closely with the underlying monitoring data, limiting its utility in improving predictions. The information captured by service descriptions may already encompass much of the relevant details about the associated components, leading to redundancy when including component descriptions in the model. 
In conclusion, while the inclusion of component descriptions adds supplementary information to the predictive model, our analysis suggests that it does not substantially enhance the model's predictive performance. 
}
\\
{
Next, we  summarize the insights from the experiments:
\begin{enumerate}
\item Additional contextual information, such as service description, aids in better classification of the SLO and resource classes.

\item The effect of additional service context is more prominent on SLO classes. This is intuitive, as the overall description helps to provide better context on the service level objectives. As resource classes deal with the underlying resources within each service, low-level granular information, such as service code snippets, may be able to provide relevant context.

\item Component description does not provide relevant context to the SLO and relevant classes. This is intuitive, as monitors are often used to monitor service level features and guarantees. Therefore, component description might become irrelevant.
\end{enumerate}
It is interesting to note that additional irrelevant information sometimes tends to reduce the performance. Additional contextual information should align with the task to achieve better performance.
}



\section{Lessons Learned and Threats} \label{discussion}
\paragraph{\textbf{Does additional contextual data help to boost performance?}}
The experiment results from the root cause analysis and monitor management experiments clearly indicate enhanced performance with the addition of contextual data. 
For root cause analysis of dependency failures, including the upstream service information along with the in-context historical examples yield the best results. The recommendations from this prompt are more readable and concise, as confirmed during Human Evaluation. Additionally, summarizing the service descriptions further improved the model output. 
The experiment results from monitor management suggest that additional contextual data, particularly from service descriptions, can indeed improve performance in monitor classification tasks, with a more significant impact observed for SLO Classes as compared to resource classes. 

\paragraph{\textbf{What additional contextual data are useful?}}
We can leverage the nature of the task to identify the additional contextual information needed.
For instance, it is intuitive that service descriptions can help in better predicting the service level objectives associated with a monitor consistently, while information on the underlying service code might help to better identify the resource characteristics within a service and, consequently, the resource classes.
Similarly, for root cause analysis, OCEs usually refer to upstream dependencies and their service property to understand and localize critical failures. For other types of failures, we might need to retrieve additional contextual information such as source codes or logs.

\paragraph{\textbf{\emph{Threats to Validity.}}}
There are several threats to our study. We evaluated the performance on a subset of incidents and monitor data within \CompanyX. The performance of predicted root cause accuracy may vary if evaluated on a different dataset from other services within \CompanyX. While the semantic and lexical metrics are widely used for evaluation of natural language recommendations, these metrics are computed using pre-trained models. Therefore, the results might vary if we train these metric computation models using incident dataset. Moreover, the human evaluation is done on a small sample of incidents, as it is challenging to scale these time-consuming manual process. For monitor management experiments, as different organizations might follow different configuration and processes, the classification accuracy might vary if evaluated on data from other organization or even other services from \CompanyX which were not included in our study.


\section{Related Work} \label{related}
\subsection{LLMs in software engineering}
Motivated by recent advancement and success, LLMs have been widely adopted in systems and software engineering domain. Github copilot is a popular example where GPT models are being used for automatic code generation from natural language text. LLMs are also leveraged for solving various other tasks including code generation~\cite{chen2021evaluating,xu2022systematic}, docstring generation~\cite{chen2021evaluating,ahmed2022few}, code repair~\cite{fan2022improving,joshi2023repair}, program analysis and synthesis~\cite{jain2022jigsaw,nijkamp2022codegen} and software testing~\cite{wang2023software}.
Jain \etal~\cite{jain2022jigsaw} propose to augment LLMs using program analysis and synthesis technique to achieve better performance. 
LANCE~\cite{mastropaolo2022using} utilizes fine-tuned T5 to automatically generate logs for Java methods. Similarly, VulRepair~\cite{fu2022vulrepair} tool utilizes a fine-tuned T5 model to automatically suggest vulnerability fixes. In contrast to these studies, we propose to leverage state-of-the-art LLM models for automating parts of incident management lifecycle by unitizing contextual data from different stages of SDLC. 

\subsection{Incident Management}
Incident management emerges as a popular research area in systems and software engineering due to its practical implication across large cloud systems. Several empirical studies have been conducted to understand and classify production outages, either delving into the types of issues that caused incidents~\cite{alquraan2018analysis,gao2018empirical,zhang2021understanding} or issues pertinent to a particular service or system~\cite{yuan2014simple,liu2019bugs,ghosh2022fight}. Several machine learning models have also been proposed to automate parts of incident management lifecycle, such as triaging~\cite{azad2022picking,chen2019continuous}, incident linking~\cite{chen2020identifying,ghosh2024dilink}, diagnosis~\cite{bansal2020decaf,luo2014correlating}, mitigation~\cite{jiang2020mitigate}. Ahmed \emph{et al.}~\cite{ahmed2023recommending} first proposed to finetune a LLM model for automatically generating the root cause and mitigation recommendations. Chen \emph{et al.}~\cite{chen2023empowering} proposed to use in-context examples with retrieval augmented generation (RAG) framework for categorization of root causes for a specific type of service. Zhang \emph{et al.}~\cite{zhang2024automated} recently demonstrate that augmenting in-context examples help in improving the root cause recommendation quality. However, these works only leverage incident metadata for finetuning and retrieval purpose. In contrast, we demonstrate at scale that augmenting additional contextual data such as service and dependency information along with in-context examples can significantly boost the accuracy of recommendations. 

\subsection{Monitor Management}
\cite{surianarayanan2019cloud,montes2013gmone,aceto2013cloud} provide an overview of cloud monitoring and describe the characteristics of a good monitoring system. They discuss the stages of monitoring, including data collection, analysis, and decision-making, and list several commercial and open-source tools available for monitoring, such as VMware Hyperic, Virtualization Manager of SolarWinds, and Azure Monitor. Nair \etal~\cite{nair2015learning} presents a hierarchical monitoring framework for Microsoft's internal distributed data storage and batch computing service, which applies machine learning algorithms for issue detection. 
\cite{mogul2017thinking,mogul2019nines,jayathilaka2015response} discusses the challenges of defining SLOs for cloud services and highlights the difficulties cloud providers face in meeting the reliable and adequate performance promises that cloud customers want, especially in the face of arbitrary customer behavior and hidden interactions.
\cite{ding2019characterizing,ding2020characterizing} address common misconceptions and practical challenges of defining and managing SLOs, especially implementing effective alerting systems that are cognizant of the desired level of granularity and service dependencies. 
These works assume that the ``what'' resource to monitor and ``which'' metric to monitor are already established, and then develop techniques that can help in faster incident detection or alert refinement. Once the fundamental problem of monitor structure and design is solved, these approaches can be augmented to the proposed framework.

\section{Conclusion} \label{conclusion}
In this work, we demonstrate the importance of augmenting the LLM prompt with data from different stages of software development lifecycle while solving two critically important tasks in incident management with real-world dataset from \CompanyX: (1) while generating root cause recommendations for dependency failure related incidents, including contextual information such as upstream service dependencies and their properties, can significantly boost the accuracy of recommendations; and (2) while identifying resource and SLO classes for an existing monitor, leveraging service architecture and properties along with monitor metadata improves the classification accuracy. We expect that this work will motivate other studies that leverage LLMs with X-lifecycle data for automating parts of incident management lifecycle.

\bibliographystyle{ACM-Reference-Format}
 \bibliography{Reference}
\end{document}